\documentclass{ws-rv9x6}

\begin{document}

\setcounter{chapter}{0}

\chapter{WAS EINSTEIN RIGHT? TESTING RELATIVITY AT THE CENTENARY}

\markboth{C. M. Will}{Was Einstein Right?}

\author{Clifford M. Will}

\address{Department of Physics and McDonnell Center for the Space
Sciences, \\
Washington University, St. Louis MO 63130 USA \\
E-mail: cmw@wuphys.wustl.edu
}

\begin{abstract}
We review the experimental evidence for Einstein's special and 
general relativity.  A variety of high precision null experiments
verify the weak equivalence principle and local Lorentz invariance,
while gravitational redshift and other clock experiments support local
position invariance.  Together these results confirm the
Einstein Equivalence Principle which underlies the concept that
gravitation is synonymous with spacetime geometry, and must be
described by a metric theory.
Solar system experiments that test the weak-field, post-Newtonian
limit of metric theories strongly favor
general relativity.  The Binary
Pulsar provides tests of gravitational-wave damping
and of strong-field general relativity.  Recently discovered binary
pulsar systems may provide additional tests.  Future and ongoing
experiments, such as the 
Gravity Probe B Gyroscope
Experiment, satellite tests of the Equivalence principle, and tests of
gravity at short distance to look for extra spatial dimensions could 
constrain extensions of general relativity.
Laser interferometric
gravitational-wave observatories on Earth and in space 
may provide new tests
of gravitational theory via detailed measurements of  
the properties of gravitational waves. 
\end{abstract}

\section{Introduction}
\label{sec:intro}

When I was a first-term graduate student some 36 years ago, it was said
that the field of general relativity is ``a theorist's paradise
and an experimentalist's purgatory''.  To be sure, there were some
experiments: Irwin Shapiro, then at MIT, had just measured the
relativistic
retardation of radar waves passing the Sun (an effect that now bears
his name), Robert Dicke of Princeton was claiming that the
Sun was flattened in an amount that would mess up
general relativity's success with Mercury's perihelion advance,
and Joseph Weber
of the University of Maryland was just about to announce (40 years
prematurely,
as we now know) the detection of gravitational waves.  Nevertheless
the field was dominated by theory and by theorists.  The field {\em circa}
1970
seemed to reflect Einstein's own attitudes: although he was not
ignorant of experiment, and indeed had a keen insight into the
workings of the physical world, he felt that the bottom line was the
{\em theory}.  As he once famously said, if experiment were to contradict
the theory, he would have ``felt sorry for the dear Lord''.

Since that time the field has been completely transformed, and today
at the centenary of Einstein's {\em annus mirabilis}, experiment is a
central, and in some ways dominant component of gravitational
physics.  I know no better way to illustrate this than to cite the
first regular article  of the 15 June 2004 issue
of Physical Review D:  the author list of this ``general relativity''
paper fills an entire page, and the institution list fills most of another.
This was one of the papers reporting results from the first science
run of the LIGO laser interferometer
gravitational-wave observatories, but it brings to mind papers in
high-energy physics, not general relativity!
The breadth of current experiments, ranging from tests of classic general
relativistic effects such as the light bending and the Shapiro delay,
to searches for short-range violations of the inverse-square law, to
the operation of a space experiment to measure the relativistic
precession of gyroscopes, to the construction and operation
of gravitational-wave detectors, attest to the ongoing vigor of
experimental gravitation.    

Because of its elegance and simplicity, and because of its empirical
success, general relativity has become the foundation for our understanding
of the gravitational interaction.  Yet modern developments in particle
theory suggest that it is probably not the entire story, and that 
modification of the basic theory may be required at some level.  String
theory generally predicts a proliferation of scalar fields 
that could result in alterations of general relativity reminiscent of the
Brans-Dicke theory of the 1960s.  In the presence of extra dimensions, 
the gravity that we feel on our four-dimensional ``brane'' of a
higher dimensional world could be somewhat different from a pure
four-dimensional general relativity.  Some of these ideas have motivated the
possibility that fundamental constants may actually be dynamical variables,
and hence may vary in time or in space.  However, any theoretical
speculation along these lines must abide by the best current empirical
bounds.  Decades of high-precision tests of general relativity have produced
some very tight constraints.  
In this article I will review the
experimental situation, and assess how well, after 100 years, Einstein
got it right.

We begin in Sec. \ref{sec:eep} 
with the ``Einstein equivalence principle'', which underlies
the idea that gravity and curved spacetime are synonymous, and
describe its empirical
support.  Section \ref{sec:solarsystem} describes solar system tests
of gravity in terms of 
experimental bounds on a set of ``parametrized post-Newtonian'' (PPN) 
parameters.  In Section
\ref{sec:pulsars} we discuss tests of general relativity using binary pulsar
systems.  Section \ref{sec:waves} describes tests of gravitational theory
that could be carried out using future observations of gravitational
radiation.  Concluding remarks are made in Section \ref{sec:conclude}.  For
further
discussion of topics in this chapter, and for references to the
literature, the reader is referred to {\it Theory and Experiment in
Gravitational Physics}\cite{tegp}
and to the ``living'' review articles\cite{livrev,stairs,mattingly}.

\section{The Einstein Equivalence Principle}
\label{sec:eep}

The Einstein equivalence principle
(EEP) is a powerful and far-reaching principle, which  states that

\begin{itemize}

\item
test bodies fall with the same acceleration independently of their
internal structure or composition (Weak Equivalence Principle, or WEP),

\item
the outcome of any local
non-gravitational experiment is independent of the velocity of
the freely-falling reference frame in which it is performed (Local
Lorentz Invariance, or LLI), and

\item
the outcome of any local non-gravitational experiment is
independent of where and when in the universe it is performed
(Local Position Invariance, or LPI).

\end{itemize}

The Einstein equivalence
principle is the heart of gravitational theory, for it
is possible to argue convincingly that if EEP is valid, then
gravitation must be described by
``metric theories of gravity'', which
state that  (i)~spacetime is endowed with a symmetric metric, (ii)~the
trajectories of freely falling bodies are geodesics of that
metric, and (iii)~in local freely falling reference frames, the
non-gravitational laws of physics are those written in the
language of special relativity.  

General relativity is a metric theory of gravity, but so are
many others, including the Brans-Dicke theory.  
In this sense, superstring theory is not metric, 
because of residual coupling of external, gravitation-like fields, to
matter.  Such external fields could be characterized as fields that do
not vanish in the vacuum state (in contrast, say, to electromagnetic
fields).  Theories in which varying non-gravitational constants are
associated with dynamical fields that couple to matter directly are also not
metric theories.

\subsection{Tests of the weak equivalence principle}
\label{sec:wep}

\begin{figure}[t]
\centerline{
\psfig{figure=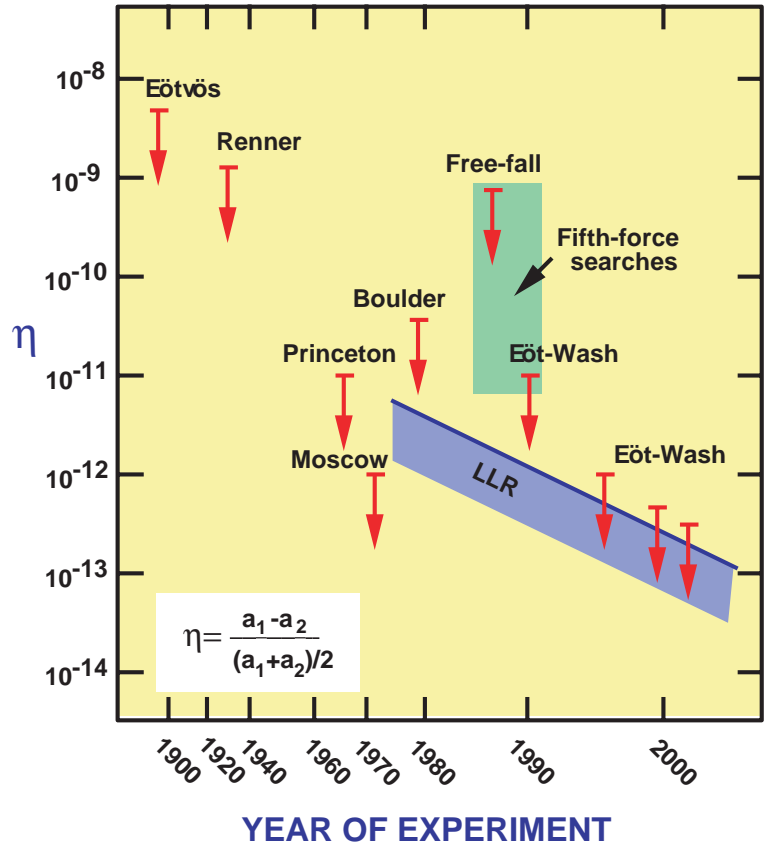,width=10cm}
}
\caption{\label{fig:wep}
Selected tests of the Weak Equivalence Principle,
showing bounds on the fractional difference in
acceleration of different materials or bodies.  ``Free-fall'' and
E\"ot-Wash experiments, along with numerous others between 1986 and
1990, were originally performed
to search for a fifth force.  Blue line and shading shows evolving 
bounds on WEP for the Earth and the Moon
from Lunar laser ranging (LLR).}
\end{figure}

To test the weak equivalence principle, one compares the acceleration of
two
laborat\-ory-sized bodies of different composition in an external
gravitational field.  
A measurement or limit
on the fractional difference in acceleration between two bodies
yields a quantity 
$ \eta \equiv {{2 | a_1  -  a_2 |} / {| a_1 +  a_2 |}}$,
called the ``E\"otv\"os ratio'', named in honor of Baron von
E\"otv\"os, the Hungarian physicist
whose experiments carried out with torsion balances at the end of the 19th
century were the first high-precision tests of WEP\cite{eotvos}.
Later classic experiments by Dicke and Braginsky\cite{dickewep,braginsky} 
improved the bounds by several orders of
magnitude.  Additional experiments were carried out during the 1980s
as part of a search for a putative ``fifth force'', that was motivated
in part by a reanalysis of E\"otv\"os' original data (the range of
bounds achieved during that period is shown schematically in the region labeled ``fifth
force'' in Figure \ref{fig:wep}).  

In a torsion balance, two bodies of different composition are
suspended at the ends of a rod that is supported by a fine wire or
fibre.  One then looks for a difference in the horizontal
accelerations of the two bodies as revealed by a slight rotation of
the rod.  The source of the horizontal gravitational force could be
the Sun, a large mass in or near the laboratory, or, as E\"otv\"os
recognized, the Earth itself.

The best limit on $\eta$ currently comes from 
the ``E\"ot-Wash'' experiments carried out at the
University of Washington, which used a sophisticated torsion balance
tray to compare the accelerations of bodies of different composition toward
the Earth, the Sun and the galaxy\cite{eotwash}.  
Another strong bound
comes from Lunar laser ranging (LLR), which checks the equality of
free fall of the Earth and Moon toward the Sun\cite{llr}.  The results from
laboratory and LLR experiments are:
\begin{equation}
\eta_{\rm E\ddot o t-Wash} < 4 \times 10^{-13} \,, \quad
\eta_{\rm LLR} < 5 \times 10^{-13} \,.
\end{equation}
In fact, by using laboratory materials whose composition mimics that of the
Earth and Moon, the E\"ot-Wash experiments\cite{eotwash}
permit one to infer an
unambiguous bound from Lunar laser ranging on the universality of
acceleration of gravitational binding energy at the level of $1.3 \times
10^{-3}$ (test of the Nordtvedt effect -- see Sec. \ref{sec:ppnbounds} and 
Table \ref{table:bounds}.)

In the future, the Apache Point Observatory for Lunar Laser-ranging
Operation (APOLLO) project, a joint effort by researchers from the
Universities of Washington, Seattle, and California, San Diego, plans
to
use enhanced laser and telescope technology, together with a good,
high-altitude site in New Mexico, to improve the Lunar laser-ranging
bound by as much as an order of magnitude\cite{apollo}.

High-precision WEP experiments,
can test superstring inspired models of scalar-tensor gravity, or
theories
with varying fundamental constants
in which weak violations of WEP can occur via non-metric couplings.
The project MICROSCOPE, designed to test WEP to a part in $10^{15}$
has
been approved by the French space agency CNES for a possible 2008
launch.  
A proposed NASA-ESA Satellite Test
of the Equivalence Principle (STEP) seeks to reach the level of $\eta <
10^{-18}$.  These experiments will compare the acceleration of
different materials moving in free-fall orbits around the Earth inside
a drag-compensated spacecraft.  Doing these experiments in space
means that the
bodies are in perpetual fall, whereas Earth-based free-fall experiments
(such as the 1987 test done at the University of
Colorado\cite{faller} indicated in Figure \ref{fig:wep}), are over in seconds,
which leads to significant measurement errors.    

Many of the high-precision, low-noise methods that were
developed for tests of WEP have been adapted to
laboratory tests of the inverse square law of
Newtonian gravitation at millimeter scales and below.  The goal of
these experiments is to
search for 
additional couplings to massive particles
or for the presence of large extra dimensions.
The challenge of these experiments is to
distinguish gravitation-like interactions from electromagnetic and
quantum
mechanical (Casimir) effects.  No deviations from
the inverse square law have been found to date at distances between $10 \,\mu
{\rm m}$ and $10 \, {\rm mm}$\cite{long99,hoyle1,hoyle2,kapitulnik,long03}.  

\subsection{Tests of local Lorentz invariance}
\label{sec:lli}

Although special relativity itself never benefited from the kind of
``crucial'' experiments, such as the perihelion advance of Mercury and
the deflection of light, that contributed so much to the initial acceptance of
general relativity and to the fame of Einstein, the steady
accumulation of experimental support, together with the successful
merger of special relativity
with quantum mechanics, led to its being accepted
by mainstream physicists by the late 1920s, ultimately to become part of
the standard toolkit of every working physicist.  
This accumulation
included 

\begin{itemize}

\item
the classic Michelson-Morley experiment and its 
descendents\cite{mm,shankland,townes,brillethall}, 
\item
the Ives-Stillwell, Rossi-Hall and
other tests of time-dilation\cite{ives,rossi,farley}, 
\item
tests of the independence of the speed of
light of the velocity of the source, using both binary X-ray stellar
sources and high-energy pions\cite{brecher,alvager},
\item
tests of the isotropy of the speed of light\cite{Champeney,riis,krisher}

\end{itemize}

In addition to these direct experiments, there was the Dirac
equation of quantum mechanics and its prediction of anti-particles and
spin; later would come the stunningly
successful
relativistic theory of quantum electrodynamics.

On this 100th anniversary of the introduction of special relativity,
one might ask ``what is there to test?''.  Special relativity has 
been so thoroughly
integrated into the fabric of modern physics that its validity is
rarely
challenged, except by cranks and crackpots.
It is ironic then, that during the past several years, a vigorous
theoretical and experimental effort has been launched, on an
international
scale, to find violations of special relativity.
The motivation for this effort is not a desire
to repudiate Einstein, but to look for
evidence of new physics ``beyond'' Einstein, such as apparent
violations
of Lorentz invariance that might result from certain models of quantum
gravity.  
Quantum gravity asserts that there is a fundamental length scale
given by the Planck length, $L_p = (\hbar G/c^3)^{1/2} = 1.6 \times
10^{-33}
\, {\rm cm}$, but since length is not an invariant quantity
(Lorentz-FitzGerald contraction), then there could be a violation of
Lorentz
invariance at some level in quantum gravity.   In brane world
scenarios, while
physics may be locally Lorentz invariant in the higher dimensional
world,
the confinement of the interactions of normal physics to our
four-dimensional ``brane'' could induce apparent Lorentz violating
effects.
And in models such as string theory, the presence of additional
scalar,
vector and tensor long-range fields that couple to matter of the
standard
model could induce effective violations of Lorentz symmetry.  
These and other ideas have motivated
a serious
reconsideration of how to test Lorentz invariance with better
precision and
in new ways.

A simple 
way of interpreting some of these experiments is to suppose that
a non-metric coupling to the 
electromagnetic interactions results in a change in the speed of
electromagnetic radiation $c$ relative to the limiting
speed of material test particles $c_0$, in other words, $c \ne c_0$.  
In units where $c_0=1$, this would result in an action for charged
particles and electromagnetic fields given, in a preferred reference
frame (presumably that of the cosmic background radiation), by
\begin{eqnarray}
 I & = & - \sum_a m_{0a} \int (1-v_a^2)^{1/2} dt
  + \sum_a e_a  \int
   (-\Phi + {\bf A} \cdot {\bf v_a}) dt
   \nonumber\\
   &&+ \frac{1}{8\pi} \int (E^2 - c^2 B^2) d^3xdt \,,
   \label{c2action}
   \end{eqnarray}
where $\Phi = -A_0$, ${\bf E} = -{\bf \nabla}\Phi - {\dot {\bf A}}$,
and
${\bf B} = {\bf \nabla} \times {\bf A}$.
This is sometimes called the ``$c^2$'' 
framework\cite{hauganwill,gabriel}; it is a special case of the
``$TH\epsilon\mu$'' framework of Lightman and Lee\cite{ll73} for
analysing non-metric theories of gravity, and of the
``standard model extension'' (SME) of Kostalecky and 
coworkers\cite{kostalecky1,kostalecky2,kostalecky3}.
Such a Lorentz-non-invariant electromagnetic
interaction would cause shifts in the energy levels of atoms and
nuclei that depend on the orientation of the quantization axis of
the state relative to our velocity in the rest-frame of the
universe, and on the
quantum numbers of the state, resulting in orientation dependence of
the fundamental frequencies of such atomic clocks.  
The magnitude of these ``clock
anisotropies'' would be proportional to 
$\delta \equiv | c^{-2}-1|$. 

\begin{figure}[t]
\centerline{
\psfig{figure=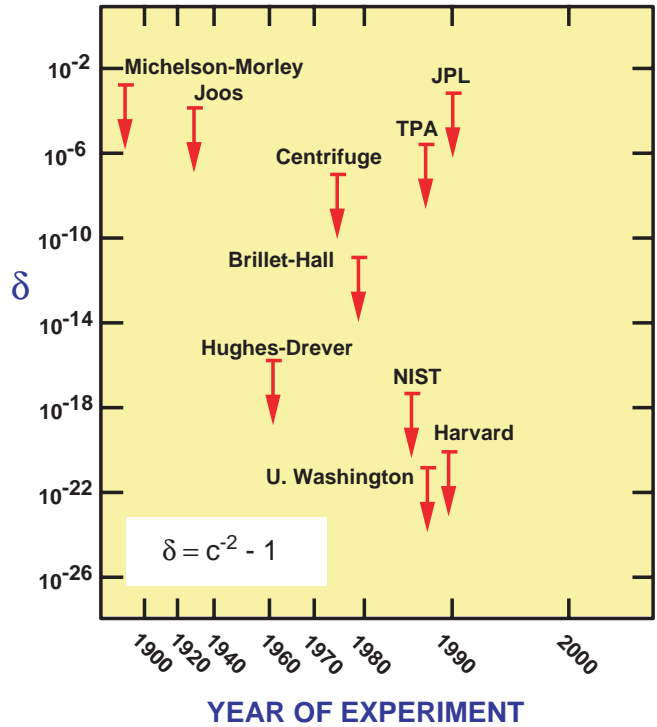,width=10cm}
}
\caption{\label{fig:lli}
Selected tests of local Lorentz invariance,
showing bounds on the parameter $\delta = c^{-2}-1$, where $c$ is the speed
of propagation of electromagnetic waves in a preferred reference frame, in
units in which the limiting speed of test particles is unity.}
\end{figure}

The earliest
clock
anisotropy experiments were those of Hughes and Drever, although their
original motivation was somewhat different\cite{hughes,drever}. 
Dramatic improvements were made in the 1980s using
laser-cooled trapped atoms and ions\cite{prestage,lamoreaux,chupp}.
This technique made
it possible
to reduce the broading of resonance lines caused by collisions,
leading to improved bounds on $\delta$ shown in Figure \ref{fig:lli}
(experiments
labelled NIST, U. Washington and Harvard, respectively).

The SME and other frameworks\cite{jacobson} 
have been used to analyse
many new experimental
tests of local Lorentz invariance, including comparisons of resonant
cavities with atomic clocks, and tests of dispersion and birefringence in
the propagation of high energy photons from astrophysical sources.
Other testable effects of Lorentz invariance violation include
threshold
effects in particle reactions,
gravitational Cerenkov radiation, and neutrino oscillations.
Mattingly\cite{mattingly} gives
a thorough and up-to-date review of both the theoretical
frameworks and the experimental results.

\subsection{Tests of local position invariance}
\label{sec:lpi}

Local position invariance, requires, 
among other things, that the internal binding energies
of atoms be independent of location in space and time, when measured
against some standard atom.  This means that a comparison of the rates
of two
different kinds of clocks should be independent of location or epoch,
and that the frequency shift of a signal sent between
two identical clocks at different locations is simply a consequence
of the apparent Doppler shift between a pair of inertial frames
momentarily comoving with the clocks at the moments of emission and
reception respectively.  The relevant parameter in the frequency shift
expression $\Delta f/f = (1+\alpha) \Delta U/c^2$, is 
$\alpha \equiv {{\partial \ln E_B} / {\partial (U/c^2)}} $, where $E_B$ is
the atomic or nuclear binding energy, and $U$ is the external
gravitational potential.  If LPI is valid, the binding energy should
be independent of the external potential, and hence $\alpha=0$.  
The best
bounds come from a 1976 rocket redshift experiment using Hydrogen
masers, and a 1993 clock
intercomparison experiment (a ``null'' redshift
experiment)\cite{vessot,godone,prestage2}.  The results are:
\begin{equation}
\alpha_{\rm Maser} < 2 \times 10^{-4} \,, \quad
\alpha_{\rm Null} <  10^{-4} \,.
\end{equation}
Recent ``clock comparison'' tests of LPI
were designed to look for possible variations of the fine structure
constant on a cosmological timescale.  An experiment done at 
the National Institute of Standards and
Technology (NIST) in Boulder 
compared  laser-cooled mercury ions with
neutral cesium atoms over a two-year period, while an experiment done
at the Observatory of Paris
compared laser-cooled cesium and rubidium atomic fountains over five
years;
the results showed that the fine structure constant $\alpha$ is constant
in time to a part in
$10^{15}$ per year\cite{salomon,bize}.  
Plans are being developed to perform such clock
comparisons in space, possibly on the International Space Station.

A better bound on $d\alpha/dt$ comes from
analysis of 
fission yields of the Oklo natural reactor, which occurred in Africa 2
billion years ago, namely  
$(\dot \alpha /\alpha)_{\rm Oklo} < 6 \times 
10^{-17} ~{\rm yr}^{-1}$\cite{damourdyson}.  
These and other bounds on variations of
constants, including 
reports (later disputed) of positive evidence for variations from
quasar spectra, are discussed by Martins and others in Ref.\cite{jenam}.

\section{Solar-system tests}
\label{sec:solarsystem}

\subsection{The parametrized post-Newtonian framework}
\label{sec:ppn}

It was once customary to discuss experimental tests of general
relativity in terms of the ``three classical tests'', the
gravitational redshift, which is really a test of the EEP, not of
general relativity itself (see Sec. \ref{sec:lpi}); 
the perihelion advance of Mercury, the
first success of the theory; and the deflection of light, whose
measurement in 1919 made Einstein a celebrity.  However, the
proliferation of additional tests as well as of well-motivated
alternative metric theories of gravity, made it desirable to develop a
more general theoretical framework for analysing both experiments and
theories.  

This ``parametrized post-Newtonian (PPN) framework'' dates
back
to Eddington in 1922, but was fully developed by Nordtvedt and Will in
the period 1968 - 72.
When we confine attention to metric theories of gravity, and further
focus on the slow-motion, weak-field limit appropriate to the solar
system and similar systems, it turns out that, in a broad class of metric
theories, only the numerical
values of a set of parameters vary from theory to theory.  The framework
contains ten PPN parameters: $\gamma$, related to the amount
of spatial curvature generated by mass; $\beta$, related to
the degree of non-linearity in the gravitational field; $\xi$,
$\alpha_1$, $\alpha_2$, and $\alpha_3$, which determine whether the
theory violates local position invariance or local Lorentz invariance
in {\it gravitational} experiments (violations of the Strong
Equivalence Principle); and $\zeta_1$, $\zeta_2$, $\zeta_3$ and
$\zeta_4$, which describe whether the theory 
has appropriate momentum conservation laws.  For a complete exposition
of the PPN framework see Ref. \cite{tegp}.

A number of well-known relativistic effects can be expressed in terms
of these PPN parameters:

\noindent
{\em Deflection of light:}
\begin{eqnarray}
\Delta \theta &=& \left (\frac{1+\gamma}{2} \right ) \frac{4GM}{dc^2}
\nonumber \\
&=& \left (\frac{1+\gamma}{2}  \right ) \times 1.7505 \frac{R_\odot}{d} \, {\rm arcsec}
\,, 
\end{eqnarray}
where $d$ is the distance of closest approach of a ray of light to a
body of mass $M$, and where the second line is the deflection by the
Sun, with radius $R_\odot$. 

\noindent
{\em Shapiro time delay:}
\begin{equation}
\Delta t = 
\left (\frac{1+\gamma}{2} \right ) 
\frac{4GM}{c^3} \ln \left [ \frac{(r_1+{\bf x}_1 \cdot {\bf n})
(r_2-{\bf x}_2 \cdot {\bf n})}{d^2}
\right ] \,,
\end{equation}
where 
$\Delta t$ is the excess travel time of a round-trip electromagnetic
tracking signal,
${\bf x}_1$ and ${\bf x}_2$ are the locations 
relative to the body of mass $M$ of the emitter and receiver of
the round-trip radar tracking signal ($r_1$ and $r_2$ are the
respective distances) and ${\bf n}$ is the direction of the outgoing
tracking signal.

\noindent
{\em Perihelion advance:}
\begin{eqnarray}
\frac{d \omega}{dt} &=& \left ( \frac{2 +2\gamma -\beta}{3} \right )
\frac{GM}{Pa(1-e^2)c^2} 
\nonumber\\
&=& \left (\frac{2 +2\gamma -\beta}{3} \right )\times 42.98 
\, {\rm arcsec/100 \, yr} \,,
\end{eqnarray}
where 
$P$, $a$, and $e$ are the period, 
semi-major axis and eccentricity of the planet's
orbit; the second line is the value for Mercury.

\noindent
{\em Nordtvedt effect:}
\begin{equation}
\frac{m_G - m_I}{m_I} = \left (4\beta -\gamma -3 
- \frac{10}{3} \xi 
-\alpha_1 - \frac{2}{3} \alpha_2 - \frac{2}{3} \zeta_1 - \frac{1}{3}\zeta_2 
\right ) \frac{|E_g|}{m_Ic^2} \,,
\end{equation}
where $m_G$ and $m_I$ are the gravitational and inertial masses of a
body such as the Earth or Moon, and $E_g$ is its
gravitational binding energy.   A non-zero Nordtvedt effect would
cause
the Earth and Moon to fall with a different acceleration toward the
Sun.

\noindent
{\em Precession of a gyroscope:}
\begin{eqnarray}
\Omega_{FD} &=& -\frac{1}{2} \left (1+\gamma + \frac{\alpha_1}{4} \right )
\frac{G}{r^3c^2} ({\bf J}-3{\bf n}\, {\bf n}\cdot {\bf J} )  \,,
\nonumber\\
&=& \frac{1}{2} \left (1+\gamma + \frac{\alpha_1}{4} \right ) \times
0.041 \,{\rm arcsec \, yr^{-1}} \,,
\nonumber\\
\Omega_{Geo} &=& - \frac{1}{2} (1 +2\gamma) {\bf v} \times \frac{Gm{\bf
n}}{r^2c^2} \,.
\nonumber\\
&=& \frac{1}{3} (1 +2\gamma) \times 6.6 \, {\rm arcsec \, yr^{-1}} \,,
\label{gyro}
\end{eqnarray}
where $\Omega_{FD}$ and $\Omega_{Geo}$ are the
precession angular velocities caused by the dragging of inertial
frames
(Lense-Thirring effect) and by the geodetic effect, a combination of
Thomas precession and precession induced by spatial curvature;
$J$ is the angular
momentum of the Earth, and ${\bf v}$, ${\bf n}$ and $r$ are the
velocity, direction, and distance of the
gyroscope.  The second line in each case is the corresponding value for
a gyroscope in polar Earth orbit at about 650 km altitude (Gravity
Probe B, Sec. \ref{sec:gpb}).

In general relativity, $\gamma=1$, $\beta=1$, and the
remaining parameters all vanish.  

\begin{table}[b]
\tbl{Current Limits on the PPN Parameters}
{\begin{tabular}{c l c l}
\noalign{\smallskip}
\hline
Parameter&Effect&Limit&Remarks \\
\hline
$\gamma-1$&(i) time delay&$2.3 \times 10^{-5}$&Cassini tracking \\
&(ii) light deflection&$3 \times 10^{-4}$&VLBI \\
$\beta-1$&(i) perihelion shift&$3 \times 10^{-3}$&$J_2=10^{-7}$ from\\
&&&helioseismology \\
&(ii) Nordtvedt effect&$5 \times 10^{-4}$&$\eta=4\beta-\gamma-3$
assumed \\
$\xi$&Earth tides&$10^{-3}$&gravimeter data \\
$\alpha_1$&orbital polarization&$10^{-4}$&Lunar laser ranging
\\
&&&PSR J2317+1439\\
$\alpha_2$&solar spin&$4 \times 10^{-7}$&alignment of Sun
\\
&precession&&and ecliptic \\
$\alpha_3$&pulsar acceleration&$2 \times 10^{-20}$
&pulsar $\dot P$ statistics \\
$\eta^1$&Nordtvedt effect&$10^{-3}$&Lunar laser ranging \\
$\zeta_1$&-- &$2 \times 10^{-2}$&combined PPN bounds \\
$\zeta_2$&binary motion&$4 \times 10^{-5}$&$\ddot P_p$
for PSR 1913+16 \\
$\zeta_3$&Newton's 3rd law&$10^{-8}$&Lunar acceleration \\
$\zeta_4$&-- &-- &not independent \\
\hline
\end{tabular}}
\begin{tabnote}
$^1$Here $\eta = 4\beta -\gamma -3 - 10 \xi /3 -\alpha_1
-2 \alpha_2 /3 - 2\zeta_1 /3 - \zeta_2 /3 $
\end{tabnote}
\label{table:bounds}
\end{table}

\subsection{Bounds on the PPN parameters}
\label{sec:ppnbounds}

Four decades of experiments, ranging
from the standard light-deflection and perihelion-shift tests, to
Lunar laser ranging, planetary and satellite tracking tests of the
Shapiro time delay, and
geophysical and astronomical observations, have placed bounds
on the PPN parameters that are consistent with general relativity.
The
current bounds
are summarized in Table \ref{table:bounds}.  

\begin{figure}[t]
\centerline{
\psfig{figure=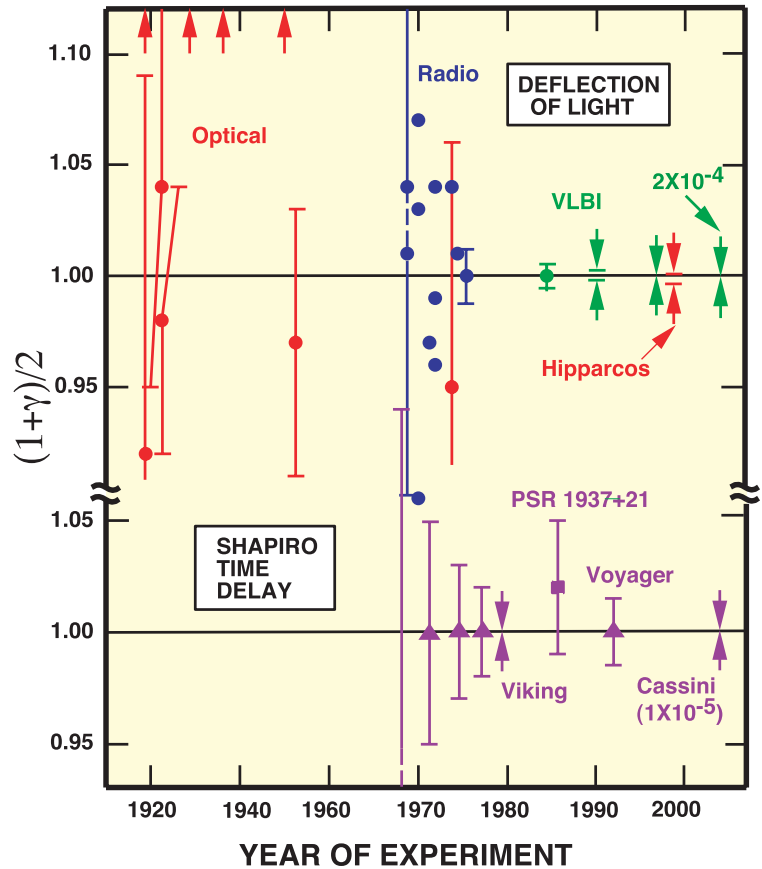,width=10cm}
}
\caption{
Measurements of the coefficient
$(1 + \gamma )/2$ from observations of the deflection of light and of
the Shapiro delay in propagation of radio signals near the Sun.
The general relativity prediction is unity.  ``Optical'' denotes
measurements of stellar deflection made during solar eclipes, ``Radio''
denotes interferometric measurements of radio-wave deflection, and ``VLBI''
denotes Very Long Baseline Radio Interferometry.
``Hipparcos'' denotes the European optical astrometry satellite.
Arrows denote
values well off the chart
from one of the 1919 eclipse expeditions and from others through
1947.  Shapiro delay
measurements using the Cassini spacecraft on its way to Saturn
yielded tests at the 0.001
percent
level, and light deflection
measurements using VLBI have reached 0.02 percent.}
\label{fig:gamma}
\end{figure}

To illustrate the dramatic progress of
experimental gravity since the dawn of Einstein's theory, 
Figure \ref{fig:gamma}
shows a history  of results for $(1+\gamma)/2$,
from the 1919 solar eclipse measurements of
Eddington and his colleagues (which made Einstein a public
celebrity), to modern-day measurements using
very-long-baseline radio
interferometry (VLBI), advanced radar tracking of spacecraft,
and orbiting astrometric satellites such as
Hipparcos. 
The most recent results include a measurement of the Shapiro delay
using the {\em Cassini} spacecraft\cite{cassini}, 
and a measurement of the bending
of
light via analysis of VLBI data on
541 quasars and compact radio galaxies distributed over the entire
sky\cite{shapiro}.

The perihelion advance of Mercury, the first of Einstein's
successes, is now known to agree with observation to a few parts in
$10^3$.  Although there was controversy during the 1960s about this
test because of Dicke's claims of an excess solar oblateness, which would
result in an unacceptably large Newtonian contribution to the
perihelion advance, it is now known from helioseismology that the
oblateness is of the order of a few parts in $10^7$, as expected from
standard solar models, and too small to affect Mercury's orbit, within the
experimental error.

Scalar-tensor theories of gravity are characterized by a coupling function
$\omega(\phi)$ whose size is inversely related to the ``strength'' of the
scalar field relative to the metric.  In the solar system, the 
parameter $|\gamma -1|$, for
example is equal to $1/(2+\omega(\phi_0))$, where $\phi_0$ is the value of
the scalar field today outside the solar system.  Solar-system experiments
(primarily the Cassini results\cite{cassini}) 
constrain $\omega(\phi_0) > 40000$.   

Proposals are being developed for advanced space
missions which will have tests of PPN  parameters as key components,
including GAIA, a high-precision astrometric telescope (successor to
Hipparcos), which could
measure light-deflection and
$\gamma$ to the $10^{-6}$ level\cite{gaia}, and the Laser Astrometric Test
of Relativity (LATOR), a mission involving laser ranging to a pair of
satellites on the far side of the Sun, which could measure 
$\gamma$ to a part in $10^{8}$, and could possibly detect second-order
effects in light propagation\cite{lator}.

\subsection{Gravity Probe B}
\label{sec:gpb}

The NASA Relativity Mission 
called Gravity Probe-B recently completed its mission to  measure the
Lense-Thirring and geodetic precessions of gyroscopes in Earth orbit\cite{gpb}.  
Launched on
April 20, 2004 for a 16-month mission, it consisted of four
spherical fused quartz rotors coated with a thin layer of
superconducting niobium, spinning at 70 - 100 Hz, in a spacecraft
containing a telescope continuously pointed toward a distant guide
star (IM Pegasi).  Superconducting current loops encircling each rotor
measure the change in direction of the rotors by detecting the change
in magnetic flux through the loop generated by the London magnetic moment
of the spinning superconducting film.  The spacecraft is in a polar
orbit at 650 km altitude.  The proper motion of the guide star
relative to the distant quasars is being measured using VLBI.  
The
primary science goal of GPB is a one-percent 
measurement of the 41 milliarcsecond per year
frame dragging or Lense-Thirring effect
caused by the rotation of the Earth; its secondary goal is to
measure to six parts in $10^5$ the larger 6.6 arcsecond per year
geodetic precession caused by space curvature [Eq. (\ref{gyro})]. 

\section{The binary pulsar}
\label{sec:pulsars}

The binary pulsar PSR 1913+16, discovered in 1974 by Joseph Taylor and
Russell Hulse,
provided important new tests of general relativity, specifically of
gravitational radiation and of strong-field gravity.  Through precise
timing of the pulsar ``clock'', the important orbital parameters of the
system could be measured with exquisite precision.  These included
non-relativistic ``Keplerian'' parameters, such as the eccentricity
$e$,
and the orbital period (at a chosen epoch) $P_b$, as well as a set of 
relativistic
``post-Keplerian'' parameters.   The first PK parameter, 
$\langle \dot{\omega} \rangle$, is  the mean rate of
advance of periastron, the analogue of Mercury's perihelion shift.
The second, denoted $\gamma^\prime$ is the effect of 
special relativistic time-dilation and the gravitational redshift
on the observed phase or arrival time 
of pulses, resulting from the pulsar's orbital
motion and the gravitational potential of its companion.
The third, $\dot{P}_b$, is the rate of decrease of the orbital period;
this is taken to be the 
result of gravitational radiation damping (apart from a small
correction due to galactic differential rotation).  Two other
parameters, 
$s$ and $r$, are related to the Shapiro time delay of the pulsar signal
if the orbital inclination is such that the signal passes in the
vicinity of the companion; $s$ is a direct measure of the orbital
inclination $\sin i$.  According to GR,
the first three post-Keplerian effects depend only on $e$ and $P_b$, which are
known, and on the two stellar masses which are unknown.  By combining
the observations of PSR 1913+16 with the GR predictions, one obtains both a
measurement of the two masses, and a test of GR, since the system is
overdetermined.  The results are\cite{weisberg}
\begin{eqnarray}
m_1 = 1.4414 \pm 0.0002 M_\odot \, &,&  \quad m_2 = 1.3867 \pm 0.0002 M_\odot
\,, \nonumber \\
{\dot P_b^{\rm GR}} / {\dot P_b^{\rm OBS}} &=& 1.0013 \pm 0.0021 \,.
\end{eqnarray}

\begin{table}[t]
\tbl{Parameters of the Binary Pulsar PSR 1913+16}
{\begin{tabular}{l l l l}
\noalign{\smallskip}
\hline
Parameter&Symbol&Value$^1$ in&Value$^1$ in\\
&&PSR1913+16&J0737-3039\\
\hline
\multicolumn{4}{l}{\bf Keplerian Parameters \hfill} \\
Eccentricity&$e$&$0.6171338(4)$&$0.087779(5)$\\
Orbital Period&$P_b$ (day)&$0.322997448930(4)$ &$0.1022525563(1)$\\
\multicolumn{4}{l}{\bf Post-Keplerian Parameters \hfill} \\
Periastron &$\langle \dot\omega \rangle\,
( ^{\rm o} {\rm
yr}^{-1} )$&$4.226595(5)$&$16.90(1)$ \\
\quad advance&&&\\
Redshift/time&$\gamma^\prime$
(ms)&$4.2919(8)$&$0.38(5)$ \\
\quad dilation&&&\cr
Orbital period &$\dot P_b \, (10^{-12} )$&$-
2.4184(9)$& \\
\quad derivative&&&\\
Shapiro delay &$s$&&$0.9995(-32,+4)$\\
\quad ($\sin i$)&\\
\hline
\end{tabular}}
\begin{tabnote}
$^1$Numbers in parentheses denote errors in last
digit.
\end{tabnote}
\label{table:binary}
\end{table}

The results also test the strong-field aspects of GR in the following
way:  the neutron stars that comprise the system have very strong
internal gravity, contributing as much as several tenths of the rest
mass of the bodies (compared to the orbital energy, which is only
$10^{-6}$ of the mass of the system).  Yet in general relativity, the
internal structure is ``effaced'' as a consequence of the Strong
Equivalence Principle (SEP), a stronger version of EEP that includes {\em
gravitationally} bound bodies and local {\em gravitational} experiments.
As a result, the orbital motion and
gravitational radiation emission depend {\em only} on the masses $m_1$ and
$m_2$, and not on their internal structure.  
By contrast, in alternative metric theories, SEP is not valid in
general, and internal-structure effects can lead to significantly
different
behavior, such as the emission of dipole gravitational radiation. 
Unfortunately, in the case of scalar-tensor theories of gravity,
because the neutron stars are so similar in PSR 1913+16 (and
in other double-neutron star binary pulsar systems), dipole radiation is
suppressed by symmetry; the best bound on the coupling
parameter $\omega(\phi_0)$ from PSR 1913+16 is in the hundreds.

However, the recent discovery of the relativistic neutron star/white dwarf
binary pulsar J1141-6545, with a 0.19 day orbital period, may ultimately
lead to a very strong bound on dipole radiation, and thence on scalar-tensor
gravity\cite{bailes,esposito}.  The remarkable ``double pulsar'' 
J0737-3039 is a binary system with two detected pulsars, in a 0.10 day
orbit seen almost edge on, 
with eccentricity $e=0.09$, and a periastron advance of $17^{\rm
o}$ per year.  A variety of novel tests of relativity, neutron star
structure, and pulsar magnetospheric physics will be possible in this
system\cite{lynescience,kramer}.
For a review of binary pulsar tests, see\cite{stairs}.

\section{Gravitational-wave tests of gravitation theory}
\label{sec:waves}

The detection of gravitational radiation by either laser
interferometers or resonant cryogenic bars will, it is widely stated,
usher in a new era of gravitational-wave
astronomy\cite{Thorne1987,Barish00}.
Furthermore, it
will yield new and interesting tests of general relativity (GR) in its
radiative regime\cite{phystoday}.

\subsection{Polarization of gravitational waves}
   
A laser-interferometric or resonant bar gravitational-wave detector 
measures the local components of a symmetric $3\times3$ tensor which
is composed of the ``electric'' components of the Riemann tensor,
$R_{0i0j}$.  These six independent components can be expressed in
terms of polarizations (modes with specific transformation properties
under null rotations).  Three are transverse to the direction of
propagation, with two representing quadrupolar deformations and one
representing an axisymmetric ``breathing'' deformation.  Three modes are
longitudinal, with one an axially symmetric 
stretching mode in the propagation direction,
and one quadrupolar mode in each of the two orthogonal planes containing the
propagation direction.  General relativity predicts only the first two
transverse quadrupolar modes, independently of the source, 
while scalar-tensor gravitational waves
can in addition contain the transverse breathing mode.  More general
metric theories predict up to the full complement of 
six modes.
A suitable array of gravitational antennas could delineate or limit
the number of modes present in a given wave.  
If
distinct evidence were found of any mode other than the two 
transverse quadrupolar modes of GR, the result would be disastrous for
GR.  On the other hand, the absence of a breathing mode would not
necessarily rule out scalar-tensor gravity, because the strength
of that mode depends on the nature of the source.  

\subsection{Speed of gravitational waves}

According to GR, in the limit in which the wavelength of gravitational
waves is small compared to the radius of curvature of the background
spacetime, the waves propagate along null geodesics of the background
spacetime, {\it i.e.} they have the same speed, $c$, as light.  In
other
theories, the speed could differ from $c$ because of coupling of
gravitation to ``background'' gravitational fields.  For example, in
some theories with a flat background metric
$\mbox{\boldmath$\eta$}$,
gravitational waves follow null geodesics of $\mbox{\boldmath$\eta$}$,
while light follows null geodesics of ${\bf g}$\cite{tegp}.
In brane-world scenarios, the apparent speed of gravitational waves
could differ from that of light if the former can propagate off the
brane into the higher dimensional ``bulk''.
Another way in which the speed of gravitational waves could differ
from $c$ is if gravitation were propagated by a massive field (a
massive graviton), in which case $v_g$ would
be given by, in a local inertial frame,
\begin{equation}
\frac{v_g}{c} = \left ( 1- \frac{m_g^2c^4}{E^2} \right )^{1/2}
\approx 1 - \frac{1}{2}\frac{c^2}{f^2\lambda_g^2} \,,
\label{eq1}
\end{equation}
where $m_g$, $E$ and $f$ are the graviton rest mass, energy and
frequency, respectively, 
and $\lambda_g = h/m_gc$ is the graviton Compton wavelength ($\lambda_g \gg
c/f$ assumed). 
An example of a theory with this property is the two-tensor
massive graviton theory of Visser\cite{visser}.

The most obvious way to test for a massive graviton is to
compare the arrival times of a gravitational wave and an
electromagnetic
wave from the same event, {\it e.g.} a supernova.
For a source at a distance $D$, the
resulting bound on the difference $|1-v_g/c|$ or on $\lambda_g$
is
\begin{eqnarray}
|1- \frac{v_g}{c}| &<& 5 \times 10^{-17} 
\left ( \frac{200 \,{\rm Mpc}}{D} \right ) 
\left ( \frac{\Delta t}{1\, {\rm s}} \right )
\,, \\
\lambda_g &>& 3 \times 10^{12}~{\rm km} 
\left ( \frac{D}{200 ~{\rm Mpc}} \frac{100 ~{\rm Hz}}{f} \right )^{1/2} 
\left (\frac{1}{f\Delta t} \right )^{1/2} \,,
\label{eq2}
\end{eqnarray}
where
$\Delta t \equiv \Delta t_a - (1+Z) \Delta t_e $ is the ``time difference'', 
where $\Delta t_a$ and $\Delta t_e$ are the differences
in arrival time and emission time, respectively, of the
two signals, and $Z$ is the redshift of the source.
In many cases, $\Delta t_e$ is unknown,
so that the best one can do is employ an upper bound on
$\Delta t_e$ based on observation or modelling.

However, there is a situation in which a bound on the graviton mass
can be set using gravitational radiation alone\cite{graviton}. 
That is the case of
the inspiralling compact binary, the final stage of evolution of
systems like the binary pulsar, in which the loss of energy to
gravitational waves has brought the binary to an inexorable 
spiral toward a final
merger.  Because the frequency of the
gravitational radiation sweeps from low frequency at the initial
moment of observation to higher frequency at the final moment, the
speed of the gravitational waves emitted will vary, from lower speeds initially
to higher speeds (closer to $c$) at the end.  This will cause a
distortion of the observed phasing of the waves and result in a
shorter than expected
overall time $\Delta t_a$ of passage of a given number of cycles.
Furthermore, through the technique of matched filtering, the
parameters of the compact binary can be measured accurately\cite{jugger}, and
thereby the 
effective emission time $\Delta t_e$ can be determined accurately.

\begin{table}
\tbl{Potentially achievable 
bounds on $\lambda_g$ from gravitational-wave observations of
inspiralling compact binaries.  }
{\begin{tabular}{cccc}
\hline
$m_1 (M_\odot)$&$m_2 (M_\odot)$&Distance (Mpc)&Bound on $\lambda_g$ (km)\\
\hline
\multicolumn{4}{l}{\bf Ground-based (LIGO/VIRGO)} \\
1.4&1.4&300&$4.6 \times 10^{12}$\\
10&10&1500&$6.0 \times 10^{12}$\\
\multicolumn{4}{l}{\bf Space-based (LISA)} \\
$10^7$&$10^7$&3000&$6.9 \times 10^{16}$\\
$10^5$&$10^5$&3000&$2.3 \times 10^{16}$\\
\hline
\end{tabular}}
\label{table:graviton}
\end{table}

A full noise analysis using proposed noise curves for the advanced
LIGO ground-based detectors, and for the proposed space-based
LISA antenna yields potentially achievable bounds that
are summarized in Table \ref{table:graviton}.  
These potential bounds can be compared
with the solid bound $\lambda_g > 2.8 \times 10^{12} \, {\rm km}$, 
derived from solar system dynamics, which limit
the presence of a Yukawa modification of Newtonian gravity of the form
$V(r) =(GM/r)\exp(-r/\lambda_g)$\cite{talmadge}, 
and with the model-depend\-ent bound 
$\lambda_g > 6\times 10^{19} \,{\rm km}$ from consideration of
galactic and cluster dynamics\cite{visser}.

\subsection{Tests of Scalar-Tensor Gravity}

Scalar-tensor theories generically predict dipole gravitational radiation,
in addition to the standard quadrupole radiation, which  
results in modifications in gravitational-radiation
back-reaction, and hence in the evolution of the phasing of gravitational
waves from inspiralling sources.
The effects are strongest for systems involving a neutron star and a
black hole.  Double neutron star systems are less
promising because the small range of masses near
$1.4~M_\odot$ with which they seem to occur 
results in suppression of dipole radiation by symmetry.
Double black-hole systems turn out to be observationally
identical in the two theories, because black holes by themselves cannot
support scalar ``hair'' of the kind present in these theories.  
Dipole radiation will be present in black-hole neutron-star systems,
however, and could be detected or bounded via matched
filtering\cite{willbd}.

Interesting bounds could be obtained using observations of
low-frequency gravitational waves by a space-based LISA-type
detector.   For example, observations of
a $1.4 M_{\odot}$ NS inspiralling to a $10^3 M_{\odot}$ BH 
with a
signal-to-noise ratio of 10 could yield a
bound on
$\omega$ between $2.1 \times 10^4$ and 
$2.1 \times 10^5$, depending on whether spins play a significant role
in the inspiral\cite{scharrewill,yuneswill,bbw}.  

\section{Conclusions}
\label{sec:conclude}

Einstein's relativistic triumph of 1905 and its follow-up in 
1915 altered the course of
science.  They were triumphs of the imagination and of theory;  
experiment played a secondary role.  In the past four decades, we have
witnessed a second triumph for Einstein, in the systematic,
high-precision
experimental verification of
his theories.  Relativity has passed every test with flying colors.
But the
work is not done.  Tests of strong-field gravity in the vicinity of
black
holes and neutron stars need to be carried out.  Gammay-ray, X-ray
and gravitational-wave astronomy will play a critical role in probing
this
largely unexplored aspect of general relativity.  

General relativity is now the ``standard model'' of gravity.  But as
in
particle physics, there may be a world beyond the standard model.
Quantum
gravity, strings and branes may lead to testable effects beyond
standard
general relativity.  Experimentalists will continue a vigorous search
for such
effects using laboratory experiments, particle accelerators, space
instrumentation and cosmological observations.  
At the centenary of relativity 
it could well be said that experimentalists have joined the theorists 
in relativistic paradise.

\bigskip
This work was supported in
part by the US National Science Foundation, Grant No. PHY 03-53180.

\end{document}